\documentclass[11pt,twocolumn]{article}

\usepackage{geometry}
\usepackage{hyperref}
\usepackage{latexsym}

\usepackage{flushend}

\usepackage{booktabs}
\usepackage{graphicx}
\usepackage{newtxtext,newtxmath}
\usepackage{natbib}

\usepackage{cleveref}
\usepackage{multirow}

\usepackage{authblk}

\begin{document}

\title{An Annotated Corpus of Webtables \\for Information Extraction Tasks}

\author[1]{Erin Macdonald\thanks{Work done while the author was affiliated with the University of Alberta.}}
\author[2]{Denilson Barbosa}

\affil[1]{Intuit \\
  10423 101 St NW, Suite 2200 \\
  Edmonton, AB, Canada}

\affil[2]{Department of Computing Science\\ University of Alberta \\ Edmonton, AB, Canada}

\affil[ ]{\url{erin_macdonald@intuit.com} and \url{denilson.barbosa@ualberta.ca}}

\date{}

\maketitle

\begin{abstract}
Information Extraction is a well-researched area of Natural Language Processing with applications in web search and question answering concerned with identifying entities and relationships between them as expressed in a given context, usually a sentence of a paragraph of running text.
Given the importance of the task, several datasets and benchmarks have been curated over the years.
However, focusing on running text alone leaves out tables which are common in many structured documents and in which pairs of entities also co-occur in context (e.g., the same row of the table).
While there are recent papers on relation extraction from tables in the literature, their experimental evaluations have been on ad-hoc datasets for the lack of a standard benchmark.
This paper helps close that gap.
We introduce an annotation framework and a dataset of 217,834 tables from Wikipedia which are annotated with 28 relations, using both classifiers and carefully designed queries over a reference knowledge graph.
Binary classifiers are then applied to the resulting dataset to remove false positives, resulting in an average annotation accuracy of 94\%.
The resulting dataset is the first of its kind to be made publicly available.
\end{abstract}

\section{Introduction}\label{intro}

We endeavored to annotate hundreds of thousands of tables with relationships between either a pair of columns or the subject of the article and a table column.
We used two different methods for this.
The first was inspired by distant supervision, a commonly used supervised algorithm for annotation discussed in \ref{rw}, which required a set of tables and a knowledge graph~\cite{Mintz:2009:DS}.
The second method was comprised of a set of carefully crafted queries to pick out relational tables, also requiring a set of tables.

To do this we needed an unannotated set of tables as well as a knowledge graph and set of relations.
We used a dump of Wikipedia tables from March 2019 for a few reasons, the most important being the trustworthiness of Wikipedia articles.
Wikipedia editors follow a set of guidelines when writing articles and new content is fact-checked regularly\footnote{\url{https://en.wikipedia.org/wiki/Wikipedia:Editing\_policy}}.
This results in a more consistent and factual dataset than, for example, tables on the web used in WebTables~\cite{Cafarella:2008:WEP:1453856.1453916}.
Wikipedia also contains information about a wide variety of topics and entity types, ensuring a diverse dataset.

Our dataset is available at \url{https://doi.org/10.7939/DVN/SHL1SL} and can be cited as
\begin{quote}
E.Macdonald and D.Barbosa. Anannotated corpus of webtables for information extraction tasks, 2019. URL \url{https://doi.org/10.7939/DVN/SHL1SL}.
\end{quote}

\section{Related Work}\label{rw}

Researchers have developed a number of datasets and benchmarks for relation extraction from running text~\cite{Riedel:2010:MRM:1889788.1889799,Hendrickx:2010:STM:1859664.1859670, mesquita-etal-2019-knowledgenet}.
Traditionally, these contain a set of sentences each with a corresponding relation label~\cite{Riedel:2010:MRM:1889788.1889799}.
Each example might also have additional information like annotated or linked entities~\cite{Riedel:2010:MRM:1889788.1889799, mesquita-etal-2019-knowledgenet}.
In addition to the New York Times~\cite{Riedel:2010:MRM:1889788.1889799} and SemEval-2010 Task 8~\cite{Hendrickx:2010:STM:1859664.1859670} datasets others include the TAC (text analysis conference) relation extraction~\cite{D17-1004}, ACE~\cite{conf/lrec/DoddingtonMPRSW04}, and KnowledgeNet~\cite{mesquita-etal-2019-knowledgenet} datasets, all of which contain thousands of instances of sentences with labeled relations.

In order to compare the results of different methods on the same dataset as accurately as possible, the same metrics for evaluation must be used.
For relation extraction this frequently means accuracy, or the $F1$ score that aggregates precision and recall~\cite{Jurafsky:2009:SLP:1214993}. Some literature presents a precision-recall curve, created by adjusting the value of $n$ to collect precision values at different recall levels~\cite{Zelenko:2003:KMR:944919.944964}, and some papers report other common metrics in Information Retrieval as well, like Mean Reciprocal Rank (MRR).

\subsection{Relation Extraction from Tables}

In recent years, some researchers have attempted to make the tables on the web more useful for search systems~\cite{Cafarella:2008:WEP:1453856.1453916,Limaye:2010:ASW:1920841.1921005,Ritze:2015:MHT:2797115.2797118,Venetis:2011:RST:2002938.2002939}.
Currently, most search engines only index the text in a web page and ignore information provided by tables~\cite{Cafarella:2008:WEP:1453856.1453916}.
This means users cannot search for the huge amount of data that is described in tables that would be tedious and repetitive to discuss explicitly in plain text~\cite{Cafarella:2008:WEP:1453856.1453916}.
Table understanding is a task parallel to information extraction where the input is a table, rather than text.

The system created by Mu\~{n}oz et al.~\cite{Munoz:2014:ULD:2556195.2556266} suggests relationships between table columns and between columns and the article subject when pairs of the same row are already related in an existing knowledge graph.
Suggestions are then filtered using a classifier by analyzing features of the article, table, columns, entities, cells and resulting triples.
The authors evaluate their work on a dump of Wikipedia's tables by manually annotating 750 of the around 37 million suggested triples using three judges.
Using only triples for which there was a unanimous agreement amongst judges, five classifiers were trained, evaluated and compared, with the best achieving close to 80\% F1 (81\% precision and 77\% recall) and producing almost eight million new triples. 

Following Mu\~{n}oz et al., a group of researchers at Roma Tre University in Italy and the University of Alberta in Canada have published three methods for relation extraction on tables~\cite{Yoones:2014:KBA,Cannaviccio:2018:LWT:3201463.3201468,Cannaviccio:2018:TAR:3178876.3186029}.
All of these methods use language models, an existing knowledge graph and an additional text corpus (Clueweb) to determine the relationships in tables.
Each entity pair in a table is scored against a model for every relation and the highest-scoring relation is selected.
The highest F1 value reported in one paper is 71\%~\cite{Cannaviccio:2018:LWT:3201463.3201468} while the most recent work reports on a different metric (MRR)~\cite{Cannaviccio:2018:TAR:3178876.3186029}.

Despite the strong similarities between the work of these two groups, including the fact both used Wikipedia tables and Freebase relations, their results are hard to compare directly as they did not use the same benchmark. For example, they did not use the same set of Freebase relations, nor did they use the same Wikipedia dump. Our goal is to help ameliorate this situation by offering a common benchmark.

\subsection{Distant Supervision}

Obtaining sufficient training or testing data has always been a longstanding challenge in Machine Learning. One accepted method to overcome that challenge in the context of relation extraction from text is that of distant supervision, introduced by Mintz et al.~\cite{Mintz:2009:DS}. The idea behind distant supervision is to leverage an existing KG to annotate sentences mentioning pairs of entities which are \textbf{known} to belong to a relation in the KG and assume those sentences as positive training data for machine learning methods for relation extraction, possibly with some filtering steps to remove obvious noise. In order to generate a large dataset of tables which could be used both for evaluating but also for training relation extraction methods we also resorted to distant supervision, as discussed below.

\section{Method}\label{method}


We chose to build our dataset using well known resources that were familiar to researchers in the area as well as representative of the task at hand. Without loss of generality we chose Wikipedia to obtain the tables and Freebase as the reference KG to obtain relations. Choosing Wikipedia is perhaps obviously a good idea as the corpus is not only easy to obtain and archived periodically, but also built within a relatively strict editorial process, leading to a fairly high quality tables and text surrounding those tables (should there be methods that exploit such texts).

\newsavebox{\RelationsTable}
\savebox{\RelationsTable}{
    \begin{tabular}{@{}l|r|r||r|r@{}}
    \toprule
    \multirow{3}{*}{\textbf{Relation}}  & \multicolumn{2}{c}{\textbf{Dist. Sup.}} & \multicolumn{2}{@{\hspace*{-2.45pt}}||c}{\textbf{Querying}} \\
    \cmidrule{2-5}
    & \textbf{Tables} & \textbf{Acc.} & \textbf{Tables} & \textbf{Acc.} \\
    \midrule
    \textit{sports\_team-player} & 21,440 & 0.83 $\rightarrow$ 0.93 & 22,102 & 0.98 \\
    \hline
    \textit{actor-film} & 22,580 & 0.90 & 28,177 & 0.98 \\
    \hline
    \textit{political\_party-politician} & 8,642 & 0.79 $\rightarrow$ 0.89 & 18,014 & 0.99 \\
    \hline
    \textit{actor-character} & 1,724 & 0.49 $\rightarrow$ 0.84 & 21,462 & 0.97  \\
    \hline
    \textit{location-contains} & 11,059 & 0.92 & 6,121 & 1.00 \\
    \hline
    \textit{football\_position-player} & 2,121 & 1.00 & 13,076 & 0.87 \\
    \hline
    \textit{musician-album} & 8,049 & 0.92 $\rightarrow$ 0.96 & 8,560 & 0.97 \\
    \hline
    \textit{person-nationality} & 8,865 & 0.58 $\rightarrow$ 0.90 & 7,002 & 1.00 \\
    \hline
    \textit{director-film} & 7,019 & 0.71 $\rightarrow$ 0.81 & 4,504 & 0.97 \\
    \hline
    \textit{award-nominee} & 6 & 0.56 $\rightarrow$ 1.00 & 5,725 & 1.00  \\
    \hline
    \textit{person-graduate} & 83 & 0.54 $\rightarrow$ 1.00 & 5,408 & 1.00  \\
    \hline
    \textit{film-language} & 2,256 & 0.89 & 2,952 & 1.00 \\
    \hline
    \textit{author-works\_written} & 588 &0.63 $\rightarrow$  0.87 & 1,712 & 0.89 \\
    \hline
    \textit{producer-film} & 668 & 0.18 $\rightarrow$ 0.46 & 964 & 0.95 \\
    \hline
    \textit{writer-film} & 1,036 & 0.17 $\rightarrow$ 0.36 & 199 & 0.80 \\
    \hline
    \textit{film-music} & 932 & 0.53 $\rightarrow$ 0.75 & 399 & 0.97 \\
    \hline
    \textit{person-profession} & 420 & 0.60 $\rightarrow$ 0.73 & 716 & 0.97 \\
    \hline
    \textit{person-parents} & 122 & 0.23 $\rightarrow$ 0.65 & 731 & 0.97 \\
    \hline
    \textit{film-country} & 474 & 0.60 $\rightarrow$ 0.82 & 380 & 1.00 \\
    \hline
    \textit{musician-origin} & 112 & 0.49 $\rightarrow$ 0.75 & 632 & 1.00  \\
    \hline
    \textit{film-production\_company} & 446 & 0.59 $\rightarrow$ 0.66 & 275 & 0.98 \\
    \hline
    \textit{person-spouse} & 305 & 0.14 $\rightarrow$ 0.60 & 90 & 1.00 \\
    \hline
    \textit{film-genre} & 8 & 0.09 $\rightarrow$ 1.00 & 203 & 0.94 \\
    \hline
    \textit{person-place\_of\_birth} & 0 & -- & 180 & 1.00 \\
    \hline
    \textit{company-industry} & 59 & 0.50 $\rightarrow$ 1.00 & 12 & 1.00  \\
    \hline
    \textit{person-place\_of\_death} & 133 & 0.05 $\rightarrow$ 0.25 & 4 & 1.00 \\
    \hline
    \textit{person-religion} & 14 & 0.70 $\rightarrow$ 1.00 & 42 & 1.00 \\
    \hline
    \textit{book-genre} & 22 & 0.79 $\rightarrow$ 1.00 & 9 & 0.89  \\
    \hline\hline
    No relation & - & - & 12,491 & - \\
    \bottomrule
  \end{tabular}
  }

To choose relations to label these tables with, we considered DBpedia, which is derived from Wikipedia and Freebase, which albeit no longer updated, has been extensively used in previous research into table understanding and also relation extraction from text~\cite{Mintz:2009:DS,Cannaviccio:2018:TAR:3178876.3186029,Heiko:2016:KGR}. For example, Mintz et al.~\citet{Mintz:2009:DS} used the 23 largest relations in Freebase at the time while \citet{Cannaviccio:2018:TAR:3178876.3186029} run their experiments on 9 relations involving entities of type person, some of which overlap with the 23 from Mintz et al. Considering that both Wikipedia and Freebase cover extensively the film domain, we also selected relations from that domain in our benchmark, resulting in 28 relations to annotate tables with, provided in \Cref{table_corpus}. 

Comparing \Cref{table_corpus} with the list of 23 largest Freebase relations used by Mintz et al. one glaring difference between the use of text and tables in Wikipedia becomes obvious. For example, the relation that we could find the most tables in Wikipedia corresponded to the team an athlete plays for, because Wikipedia has many articles written by sports experts and aficionados with such information. In contrast, that relation is not among those used in the distant supervision benchmark. Similarly, that benchmark does not include the next three relations in our benchmark in terms of the number of tables we could find: \emph{actor-film}, \emph{political\_party-politician}, and \emph{actor-character}.

\begin{table}[t]
  \scalebox{0.825}{\usebox{\RelationsTable}}
  \caption{Number of tables collected and estimated accuracy with each method. Results for column pairs and article subject-column pairs are combined.}
  \label{table_corpus}
\end{table}

We restricted our benchmark to relations involving named entities that were disambiguated (via \emph{wikilinks} in the table cells) as subject and object only. We made this choice based on two factors. First, several methods in the literature do not contemplate entity disambiguation, and could not therefore be evaluated with our benchmark. Second, despite tremendous recent progress, named entity disambiguation remains a challenge and relying on automatic methods for that step would introduce error in our dataset. We will consider relaxing this restriction in future releases of the benchmark.

\begin{table}[t]
  \small
  \centering
  \begin{tabular}{@{}l|r|r||r|r@{}}
    \toprule
    \multirow{3}{*}{\textbf{Classifier}}  & \multicolumn{2}{c}{\textbf{Col Pairs}} & \multicolumn{2}{c}{\textbf{Subj-Col Pairs}} \\
    \cmidrule{2-5}
    & \textbf{Small} & \textbf{Large} & \textbf{Small} & \textbf{Large} \\
    \midrule
    \textbf{kNN ($k=1$)} & 0.69 & 0.70 & 0.72 & 0.69 \\
    \hline
    \textbf{GNB} & 0.67 & 0.61 & 0.66 & 0.57 \\
    \hline
    \textbf{kNN ($k=2$)} & 0.61 & 0.60 & 0.62 & 0.51 \\
    \hline
    \textbf{NC} & 0.54 & 0.55 & 0.68 & 0.67 \\
    \bottomrule
  \end{tabular}
  \caption{Annotation accuracy of different classifiers using different size feature vectors.}
  \label{classifiers}
\end{table}

\subsection{Collecting Tables}

We used two different methods to obtain tables from Wikipedia tables. First, we attempted to adapt distant supervision to our task. Upon realizing some crucial shortcomings, we resorted to manually crafted queries over Freebase and DBpedia to further improve our dataset.

\paragraph*{Distant Supervision.}

We first used a form of distant supervision to annotate the Wikipedia tables.
This involved gathering a list of entity pairs $(e_1, e_2)$ related in Freebase through a relation $r$ from our set of 28 relations.
We also included entities related through analogous relations in DBpedia, another public knowledge base similar to Freebase, to increase number of pairs.

For each pair $(e_1, e_2)$ collected for a relation $r$, we searched for Wikipedia tables with $e_1$ and $e_2$ appearing in cells of the same row but different columns, $c_1$ and $c_2$ (referred to as column pairs).
We make the assumption that relation $r$ holds between $c_1$ and $c_2$ and can infer the relation for all pairs in those columns.
We also searched for tables with either $e_1$ or $e_2$ as the subject of the Wikipedia article and the remaining entity appearing in any cell of a table in that article (referred to as article subject-column pairs).
We then assume the article subject is related through $r$ to the corresponding column.

\begin{figure*}[t]
  \includegraphics[width=\textwidth]{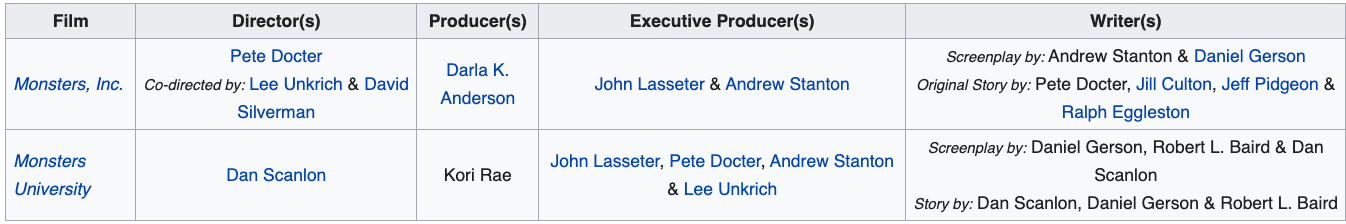}
  \caption{A table from the Wikipedia article ``Monsters Inc. (franchise)'' showing the entity ``Dan Scanlon'' in two different columns representing different relations to the film ``Monsters University''.}
  \label{mu}
\end{figure*}

These assumptions are responsible for almost all the erroneously annotated tables in this dataset.
The root of the problem is illustrated in \Cref{mu}.
Two entities can and often are related to one another through multiple relations so when we annotate a table like that in our example, we will assign both the relations director and writer between columns 1 and 2 and columns 1 and 5, two of which are incorrect.

In an attempt to mitigate these errors, we performed a small set of experiments using three different binary classifiers to mark each annotation as correct or incorrect, using the relations annotated in the previous step with the lowest accuracy.
To create training and testing data for these classifiers, we hand-annotated 200 tables for each relation, using 100 for training and the other 100 for testing.
We also use these annotations as a measure of the accuracy of the distant supervision method.
We experimented using k-nearest-neighbors, nearest centroid and Gaussian naive Bayes, building the feature vector for each annotation using a combination of the word2vec embeddings of the terms in the article title, headers, section title and section text.
We also compared how different vector lengths (generated using more or less words in those texts mentioned above) and present the results in \Cref{classifiers} where we compare the classifiers with different vector lengths on both the column pair and article subject-column pair tasks.
The results show that kNN with $k=1$ consistently out-performs other classifiers.

\paragraph{Ad-hoc queries.}

When curating the gold standard dataset for training the above classifiers, we identified some commonalities in how data for certain relations is represented in the tables.
Using this information, we composed a set of over 150 queries for each relation using column headers and section titles to run on the dataset.
There are two notable benefits to this method compared to distant supervision and one major downside.
The benefits are that it does not rely on information already present in any dataset and, when 100 tables per relation were checked manually, it proved far more accurate than distant supervision.
However the primary downside was that we annotated far fewer tables in this way.

In \Cref{table_corpus} we present an estimate of the accuracy of the tables collected for each method (before and after classifiers were applied).
We also annotated 12,491 tables with no relation present by querying for random tables in the dataset, selecting random columns and annotating those that weren't already annotated with a relation.

\section{Use Case and Benchmark Difficulty}

We briefly report in \Cref{comparison} results of a first study using the benchmark described above (citation omitted due to double blind requirements). We note that the benchmark was created \emph{prior} to conducting this study.

The baseline choses the relation in the KG that covers the most pairs of entities in the respective table, achieving very low results. Our method uses a neural network that takes into consideration not only the entities in columns but also contextual information from the article in which the table appears such as the table caption, the table headers, the title of the section in which the appears, and the first paragraph of that section. 

\begin{table}
  \centering
  \small
  \begin{tabular}{l|r|r}
    \toprule
     & \textbf{Accuracy} & \textbf{F1} \\
    \midrule
    Baseline & 0.15 & 0.27 \\
    \hline\hline
    anonymous & 0.88 & 0.95 \\
    \bottomrule
  \end{tabular}
  \caption{Accuracy and F1.}
  \label{comparison}
\end{table}

\begin{table}
  \centering
  \small
  \begin{tabular}{l|r}
    \toprule
    \textbf{Network} & \textbf{Accuracy} \\
    \midrule
    Full & 0.92 \\
    \hline
    Full - table captions & 0.91 \\
    \hline
    Full - table headers & 0.76 \\
    \hline
    Full - section paragraphs & 0.76 \\
    \hline
    Full - section titles & 0.72 \\
    \bottomrule
  \end{tabular}
  \caption{Ablation study.}
  \label{ablation}
\end{table}

\Cref{ablation} shows an ablation study indicating that among all sources of contextual information, section titles contribute the most to finding the correct relation, followed by the paragraphs and table headers. We argue these numbers, together with the low performance of the query-only baseline support the claim that our benchmark is sufficiently challenging to contribute to further development in the field.

For comparison, Mu\~{n}oz et al.~\cite{Munoz:2014:ULD:2556195.2556266} and Cannaviccio et al.~\cite{Cannaviccio:2018:TAR:3178876.3186029} report 0.71 and 0.74 F1 scores for relation extraction from Wikipedia tables. Although this comparison is imperfect, as their methods used different relations and different Wikipedia tables than one another and than us.

\section{Dataset and Code Release}

We are releasing the code used to create the dataset, the 200 manually annotated tables for each of the relations, the results of all classifiers, and the ad-hoc queries used to improve our table corpus for other researchers.

\section{Conclusion}\label{conclusion}

We draw two conclusions from this work in annotating tabular data.
The first is that a versatile method such as distant supervision is incredibly effective in discovering possible annotations for diverse types of data, but is limited by information in the knowledge graph and makes an assumption which inevitably introduces error to the dataset.
Second, we conclude that by using a combination of different methods each with their own compromises, we could build a diverse annotated dataset that is able to discover brand new relations to add to a knowledge graph.

This dataset can be very helpful in training relation extraction systems to detect relations in tables, further filtering the dataset itself to add new, accurate relations to a knowledge graph or training systems for question-answering that can return tables containing relevant information.
Future research could focus on the cleaning step applied to the tables collected with distant supervision.
Refining this method using more sophisticated classifiers or more informative feature vectors has the potential to improve the confidence in these annotations. 
Another future improvement to the benchmark would be expanding the set of relations to include those involving entities and literals (e.g., dates or other numerical values), in a way that results on the different kinds of relations could be reported separately.

\bibliographystyle{plainnat}
\bibliography{bibliography}

\end{document}